\documentclass[pra,aps,english,onecolumn,notitlepage,superscriptaddress]{revtex4-2}
\usepackage{amsmath}
\usepackage{amsfonts}
\usepackage{amssymb}
\usepackage{mathtools}
\usepackage{physics}
\usepackage{txfonts}
\usepackage[T1]{fontenc}  
\usepackage[latin1]{inputenc}
 \usepackage{babel} 
  \usepackage{epsfig,epsf,psfrag} 
\usepackage{graphicx} 
\usepackage{pslatex}
\usepackage{epic,eepic} 
\usepackage{color,pstcol}
\usepackage{pstricks} 
\usepackage{fancyhdr} 
\usepackage{tabularx}
\usepackage{url}
\usepackage{listings}
\usepackage{color}
\usepackage{centernot}
\usepackage{natbib}
\usepackage{float}
\usepackage{fancyhdr}
\begin{document}
\title{Is the essence of a quantum game captured completely in the original classical game?}
\author{Muhammed Jabir T}
\author{Nilesh Vyas}
\altaffiliation{Current affiliation:\it Telecom ParisTech, Paris, France }
\author{Colin Benjamin}
\email{colin.nano@gmail.com}
\affiliation{\it School of Physical Sciences, National Institute of Science Education \& Research, HBNI, Jatni-752050, India }
\begin{abstract}
S. J. van Enk and R. Pike in PRA 66, 024306 (2002) argue that the equilibrium solution to a quantum game isn't unique but is already present in the classical game itself. In this work, we contest this assertion by showing that a random strategy in a particular quantum (Hawk-Dove) game is unique to the quantum game. In other words, one cannot obtain the equilibrium solution of the quantum Hawk-Dove game in the classical Hawk-Dove game. Moreover, we provide an analytical solution to the quantum $2\times2$ strategic form Hawk-Dove game using randomly mixed strategies. The random strategy which we describe is Pareto optimal with their payoff classically unobtainable. We compare the quantum strategies to correlated strategies and find that the quantum Hawk-Dove game or quantum Prisoner's dilemma yields the Nash equilibrium solution.
\end{abstract}
\maketitle
\section{Introduction}
Mathematicians have been interested in parlor games since the time of Plato. It was Euler who started the new field of graph theory with a game-theoretic notion- what would be the smallest path through the seven rivers of Konigsberg \cite{1}. However, it was left to John von Neumann to put parlor games in the mathematical language under the guise of game theory\cite{vonneumann}. Game theory, when it started, was a remarkable concept which enabled economists, social scientists, statistical physicists to propose game-theoretic solutions to economic, social, and statistical physics problems \cite{3,4,5}. Quantum theory was ante-natal to game theory. However, quantum physicists were late in employing game theoretic techniques in quantum problems. It was not until quantum information theory came into being that quantum game-theoretic issues came into vogue\cite{6,7}. However, a rude blow was struck on quantum game theory by the work of van Enk and Pike \cite{8}.  They purported to show that quantum games do not have anything more to offer than classical games, albeit with the caveat that they might be relevant to quantum algorithms. This work led to an embargo in certain journals\cite{9} regarding game theory papers. van Enk and Pike's argument is based on Prisoner's dilemma, both the classical and the quantum version\cite{6}. This work aims to provide an alternative to van Enk-Pike's case study on Prisoner's dilemma. We show that van Enk-Pike's assertion is incorrect and also not universal. We offer an alternative game- the quantum Hawk-Dove game, which gives a different solution than the classical equivalent, but the latter can never capture its essence.
\par This paper is organized as follows- section II reviews the concept of quantum games, where we briefly describe Nash equilibrium and Pareto optimality using Prisoner's dilemma as an example. We elucidate van Enk-Pike's criterion and the motivation behind this work in subsequent subsections.  Section III introduces and solves the quantum Hawk-Dove game. We show how a random strategy in a quantum Hawk-Dove game gives an equilibrium solution that cannot be replicated in the classical Hawk-Dove game, unlike the case of quantum Prisoner's dilemma. We thus refute van Enk-Pike's assertion that the equilibrium solution of quantum games can be found in the classical game itself. We also introduce correlated equilibrium and contrast it with the Nash equilibrium solution. In section IV, with a brief discussion on our results, and finally, we implement the random strategy on quantum Prisoner's dilemma in the Appendix. Herein, too van Enk-Pike's assertion that the essence of quantized Prisoner's dilemma is contained in the original classical game doesn't hold up.
\section{Quantum Games}
Any situation~\cite{6} wherein a quantum system steered by two or more parties, results in the quantification of utilities, can be formulated as a quantum game. In general, a game $G$ in the strategic form is a triplet\cite{3}-
\begin{equation}
G = (N,(S_{i})_{i\in N},(u_{i})_{i\in N}),
\label{Game}
\end{equation} in Eq.~(\ref{Game}), $N= \{ 1,2,...,k\}$ defines a finite set of players, $S_{i}$ is the set of strategies of player $i$, for every player $i\in N$ and $u_{i}: S_{1}\times S_{2}\times ... \times S_{k} \rightarrow R$ ($R$ is the real space) is a function associating each vector of strategies $s = (s_{i})_{i\in N}$ with the payoff $u_{i}(s)$ of player $i$, for every player $i\in N$. For a two player strategic game, it is convenient to write Eq.~(\ref{Game}) in bimatrix form-\\
\begin{equation}
\begin{tabular}{|c|c|c|c|}
\hline
\multicolumn{2}{|c}{}& \multicolumn{2}{|c|} {\textbf{Player2}} \\
\hline
\multicolumn{2}{|c|}{} & $s_{2}^{1}$ & $s_{2}^{2}$ \\
\hline
& $s_{1}^{1}$ & $(a_{11},b_{11})$ &$(a_{12},b_{12})$\\
\textbf{Player1} & $s_{1}^{2}$ & $(a_{21},b_{21})$&$(a_{22},b_{22})$\\
\hline
\end{tabular}
\label{asd}
\end{equation}
In Eq.~(\ref{asd}), $s_{l}^{m}$ represents strategy profile with $l\in\{1,2\}$ being the player's profile and $m\in\{ 1,2\}$ the strategy for each player. A pair $(a_{ij},b_{ij})\in R^{2}$ represents the payoffs for the player 1 and 2, respectively. A two player quantum game $(\mathcal{T})$ can be enumerated as $\mathcal{T} = \{\mathcal{H}, \rho,S_{A},S_{B},P_{A}, P_{B}\}$, with $\mathcal{H}$ describing the Hilbert space of the physical system, $\rho$ the initial state and the sets $S_{A}$ and $S_{B}$ are the permissible quantum operations of the two players. $P_{A}$ and $P_{B}$ are the utility functionals or payoff operators which specify the utility for each player and operations $s_{A} \in S_{A}$, $s_{B} \in S_{B}$ are the strategies for each player.

\subsection{Nash Equilibrium and Pareto Optimality}
One of the most widely used methods for predicting the outcome of a strategic-form game is Nash equilibrium \cite{3,10}. It is a
strategy profile from which no player has a profitable deviation. A strategy vector $s^{*}= (s_{1}^{*},s_{2}^{*},\cdots,s_{n}^{*})$ is a Nash equilibrium if for each player $i \in N$ and each strategy $s_{i} \in S_{i}$ the following is satisfied:
\begin{equation}
u_{i}(s^{*}) \geq u_{i}(s_{i},s_{-i}^{*}).
\end{equation}

\par Nash equilibrium exists when each {player's} strategy is their best response to the predicted strategy of opponents. When such an equilibrium exists, no player has an incentive to deviate from it unilaterally. The idea of Nash equilibrium is also intimately connected to the concept of Pareto optimality. Suppose two or more players follow strategies to the extent that no other combination of strategies can increase at least one {person's} payoff without reducing the payoffs of others. This strategy pair is Pareto optimal. In other words, an outcome is Pareto optimal if no other outcome can help some players without adversely affecting others. But in many cases, it is not guaranteed that a Nash Equilibrium is also Pareto optimal. The most famous example where Nash equilibrium and Pareto optimality differ is the Prisoner's dilemma.

\subsubsection{Prisoner's Dilemma} In the traditional version of Prisoner's dilemma, the police have arrested two suspects (Alice and Bob) and are interrogating them in separate rooms. Each has two choices: (i) to cooperate with each other and not confess the crime ($C$), and (ii) to defect to the police and confess the crime ($D$). Payoff matrix for this game reads:
\begin{equation}
\begin{tabular}{|c|c|c|c|}
\hline
\multicolumn{2}{|c}{}& \multicolumn{2}{|c|} {\textbf{Bob}} \\
\hline
\multicolumn{2}{|c|}{} & $C$ & $D$ \\
\hline
& $C$ & $(3,3)$ &$(0,5)$\\
\textbf{Alice} & $D$ & $(5,0)$&$(1,1)$\\
\hline
\end{tabular}
\label{PD}
\end{equation}
The payoff in the above table can be explained as follows- '3' implies one year in jail while '1' implies ten years in jail. '0' represents a life sentence, while '5' implies no jail time. No matter what the other suspect does, each can improve their position by defecting to the police. If the other defects, one better do the same to avoid harsh punishment (payoff "0" or life sentence). If the other cooperates, then one can obtain the favorable treatment accorded a {state's} witness by defecting to the police (a payoff of "5" implying no jail term). Thus, defecting is the dominant strategy for each. But when both defect, the outcome is worse for both than when they both cooperate. Thus cooperate, i.e., $(C, C)$ is a Pareto optimal strategy. The aforesaid is the description of pure strategies in classical Prisoner's dilemma. Let's find out what happens if both play a mixed strategy. Consider repeated play of the game in which $p$($q$) are probabilities with which $C$ is played by Alice(Bob). Strategy $D$ is then played with probability $(1 - p)$ by Alice, and with probability $(1 - q)$ by Bob, and the player's payoff relations read
\begin{equation}
\pi_{A,B}(p,q) =
\begin{pmatrix}
p\\
1-p
\end{pmatrix}^{T}
\begin{pmatrix}
(3,3) & (0,5)\\
(5,0) & (1,1)
\end{pmatrix}
\begin{pmatrix}
q\\
1-q
\end{pmatrix}.
\end{equation}
The strategic pair $(p^{*}, q^{*})$ is Nash equilibrium when
\begin{equation}
\pi_{A}(p^{*}, q^{*}) - \pi_{A}(p, q)\geq 0, \hspace{0.5cm}\pi_{B}(p^{*}, q^{*}) - \pi_{B}(p, q)\geq 0.
\end{equation}
For the payoff matrix (\ref{PD}) these inequalities generates a strategic pair $(p^{*}, q^{*})$ = $(0,0)$ i.e, $(D,D)$ is the Nash equilibrium for mixed strategies too \cite{3,10}.

\subsection{van Enk-Pike's criterion}
S. J. van Enk and R. Pike in \cite{8} compare quantum games to classical games as regards their utility to quantum information processing. Firstly, they question the extent to which the solution of a quantum game is available in the underlying classical game itself. They argue that even though the quantum game does not solve the underlying classical game in a quantum game scenario, the equilibrium quantum solution already exists in the original classical game. Secondly, they argue that introducing an entangled state allows players to use correlations present in such a state. Hence, violating the spirit of non-cooperative games.
\par In order to illustrate this, they consider a two-player Prisoner's dilemma game between Alice and Bob denoted by the payoff matrix (\ref{PD}). The dilemma is that, $(D, D)$ is the dominant equilibrium or Nash equilibrium, but both players would prefer $(C, C)$, i.e., it is Pareto optimal. The general quantization procedure as suggested in \cite{6} would yield a quantum payoff table:

\begin{equation}
\begin{tabular}{|c|c|c|c|c|}
\hline
\multicolumn{2}{|c}{}& \multicolumn{3}{|c|} {\textbf{Bob}} \\
\hline
\multicolumn{2}{|c|}{} & $C$ & $D$ & $Q$ \\
\hline
& $C$ & (3,3) &(0,5)&(1,1)\\
\textbf{Alice} & $D$ & (5,0) &(1,1) &(0,5)\\
& $Q$ & (1,1) & (5,0)&(3,3)\\
\hline
\end{tabular}
\end{equation}

where $Q,$ is a quantum strategy defined as $ Q = iZ$ and played on a quantum game with a maximally entangled state. $(Q, Q)$ is the solution to the quantum Prisoner's dilemma game wherein $C=I$ and $D=X$. According to the van Enk-Pike criterion, the quantum solution obtained by considering the quantum strategies over an entangled state is not unique but also possible in the classical game. It can be seen from the payoff of $(3,3)$ for the $(C, C)$ strategies. Thus, the equilibrium solution obtained in the quantum game can also be secured in the classical game. Not only this one can also see $(C,Q) \& (D,D), (D,Q) \& (C,D) \mbox{ and } (Q,D) \& (D,C)$ yield exactly identical payoff's. Thus, they conclude that the essence of quantized Prisoner's dilemma is captured completely by the classical Prisoner's dilemma game.
\subsection{Motivation}
\label{motiv}
We want to challenge this assertion of van Enk-Pike, especially the fact that the equilibrium solution of quantum games is present in the original classical game itself. It is substantiated by playing a completely random strategy over an entangled quantum state using the two-player Hawk Dove game as an example, delineated in the next section. We show that playing a random strategy over a maximally entangled state yields a solution that is impossible in the original classical game. The main take-home message of our work is elucidated in the payoff table (\ref{Qhawkdove}). In this way, we negate van Enk-Pike's assertion that solutions to quantum games are available in the classical game itself. In our case, the quantum game- "playing random strategies in the non-locally entangled game of Hawk-Dove" cannot be replicated, nor its essence captured in the classical Hawk-Dove game.
\par Secondly, van Enk-Pike's comment regarding non-local correlations violating the spirit of a non-cooperative game is fallacious. There is no stopping classical correlation being present in the classical Prisoner's dilemma game. The two prisoners could be two brothers or husband-wife, but that does not change the definition of the classical Prisoner's dilemma game. In the quantum version, too, the introduction of quantum entanglement does not change the game. The two parties are not aware of the fact that the entangled state is distributed between them. The introduction of entanglement induces correlations non locally. Finally, van Enk-Pike comments comparing the solution of the factorization problem via quantum Shor's algorithm to quantum game theory. In their own words-"no classical solution for the game of efficiently factoring large numbers is known, so quantum mechanics provides a genuinely novel solution.`` We completely debunk this argument via our example of playing random strategies in the quantum Hawk-Dove game and show that it provides a unique equilibrium solution absent from the classical game. In the next section, we solve the quantum Hawk-Dove game in detail, followed by a discussion.
\section{Quantum Hawk-Dove Game}
In our work, we negate van Enk-Pike's assertion on quantum games by showing that a completely random strategy can solve the quantum game by yielding a better and unique equilibrium solution when being operated over a maximally entangled state. To show that, we consider a two-player strategic form Hawk-Dove game with complete information\cite{3}. Hawks are aggressive and always fight to take possession of a resource. These fights are brutal, and the loser is the one who first sustains the injury. The winner takes sole possession of the resource. However, Doves never fight for the resource, displaying patience and, if attacked, immediately withdraw to avoid injury. Thus, Doves will always lose a conflict against the Hawks but without sustaining any injury. If two Doves face each other, there will be a period of displaying patience with some cost (time or energy for display) to both but without any injury. It is assumed that both the Doves are equally good in displaying and waiting for a random time. In a Dove-Dove contest, both have an equal probability of winning. The winner would be the one with more patience. The classical payoff matrix is represented as follows:
\begin{equation}
\begin{tabular}{|c|c|c|c|}
\hline
\multicolumn{2}{|c}{}& \multicolumn{2}{|c|} {\textbf{Bob}} \\
\hline
\multicolumn{2}{|c|}{} & $H$ & $D$ \\
\hline
& $H$ & $\Big(\frac{v}{2}-\frac{i}{2},\frac{v}{2}-\frac{i}{2}\Big)$ & ($v$,0)\\
\textbf{Alice} & $D$ & (0,$v$) & $(\frac{v}{2}-d,\frac{v}{2}-d)$\\
\hline
\end{tabular}
\label{HD}
\end{equation}
where $v$ and $i$ are the value of resource and cost of injury, respectively. The cost of displaying patience and waiting is $d$.
Let $v= 50$, $i=100$ and $d=10$. The reason for taking these particular values is because when both players choose Hawk they will suffer a loss in the form of injury. The injury reduces the player's ability to gain the resource in the future. Thus, the injury tends to preclude gain in the future and is therefore taken to be large. On the other hand, the cost of displaying patience is kept purposely low, and in general, less than the cost of the resource. This is because one can retreat when the other chooses to be a Hawk\cite{reasonhawk}. For the aforesaid set of values the payoff table when both Alice and Bob pursue pure strategies is given as-
\begin{equation}
\begin{tabular}{|c|c|c|c|}
\hline
\multicolumn{2}{|c}{}& \multicolumn{2}{|c|} {\textbf{Bob}} \\
\hline
\multicolumn{2}{|c|}{} & $H$ & $D$ \\
\hline
& $H$ & (-25,-25) & (50,0)\\
\textbf{Alice} & $D$ & (0,50) & (15,15)\\
\hline
\end{tabular}
\label{HDclass}
\end{equation}
The Nash equilibrium for the classical Hawk-Dove game is $(H,D)$ and $(D,H)$. And $(D,D)$ is the Pareto optimal. \\

\subsection{Mixed strategy Nash equilibrium in classical Hawk-Dove Game}
Since the Hawk-Dove game gives two pure equilibrium solutions, there must exist a third Nash equilibrium\cite{3}, i.e., mixed strategy Nash equilibrium.

{Consider Alice playing Hawk with probability $p$ and Dove with $1-p$, and Bob playing his strategies with $q$ and $1-q$. Then the expected payoffs for Alice ($\$_A$) and Bob ($\$_B$) from payoff-matrix Eq.(\ref{HDclass}) are,
\begin{align}
\$_A(p,q) = -25pq + 50p(1-q) + 15(1-p)(1-q)\\
\$_B(p,q) = -25pq + 50(1-p)q + 15(1-p)(1-q)
\end{align}
Suppose (p, q) is a mixed strategy Nash equilibrium with $0 < p < 1$ and $0 < q < 1$. Then the definition of mixed strategy Nash equilibrium implies that
\begin{equation}
\frac{\partial\$_A}{\partial p}(p,q) = 0 \quad \text{and} \quad \frac{\partial\$_B}{\partial q}(p,q) = 0
\end{equation}
Therefore,
\begin{align}
\frac{\partial\$_A}{\partial p}(p,q) = -25q + 50(1-q) -15(1-q) = 0 \implies q = \frac{7}{12} \quad \text{and} \quad 1-q = \frac{5}{12}\\
\frac{\partial\$_B}{\partial q}(p,q) = -25p + 50(1-p) - 15(1-p)= 0 \implies p = \frac{7}{12} \quad \text{and} \quad 1-p = \frac{5}{12}
\end{align}
}
Thus, the probability of playing mixed strategy Hawk/Dove game leads to Nash equilibrium is when ($p$, $q$) = (7/12, 7/12). The expanded payoff matrix with the mixed strategy equilibrium $"M_{HD}"$ (where, $p = 7/12$ and $q = 7/12$) will be:
\begin{equation}
\begin{tabular}{|c|c|c|c|c|}
\hline
\multicolumn{2}{|c}{}& \multicolumn{3}{|c|} {\textbf{Bob}} \\
\hline
\multicolumn{2}{|c|}{} & $H$ & $D$ & $M_{HD}$ \\
\hline
& $H$ & (-25,-25) &(50,0)&(6.25, -14.58)\\
\textbf{Alice} & $D$ & (0,50) &(15,15) &(6.25, 35.42)\\
& $M_{HD}$ & (-14.58, 6.25) & (35.42, 6.25)&(6.25, 6.25)\\
\hline
\end{tabular}
\label{mHD}
\end{equation}
The Nash equilibrium are thus $(H,D)$ and $(D,H)$ and $(M_{HD}, M_{HD})$. One must also notice that the mixed strategy Nash equilibrium $(6.25, 6.25)$ is much lesser than Pareto optimality payoffs $(15,15)$.

{
\subsection{Correlated equilibrium in classical Hawk-Dove Game}

Mathematician Robert Aumann first discussed the concept of correlated equilibrium in his work\cite{21}. It was introduced to capture the strategic correlation opportunities that the players face once they consider the extraneous environment in which they interact. A correlated equilibrium is a solution concept that is more general than the Nash equilibrium.

Let $S:= S_{1}\times S_{2}\times ... \times S_{n}$ be the set of all strategy profiles of an n-player game. We consider a probability distribution $q$ on $S$ (i.e., $q \geq 0$, and $\sum_{s\in S}q_s = 1$). Then $q$ is a correlated equilibrium, if $\forall \, t_i \in S_i$, and for each player $i$, such that
\begin{equation}
\sum_{s\in S}q_s u_i(s) \geq \sum_{s\in S}q_s u_i(s_{-i}, t_i),
\label{corr eq}
\end{equation}
where, $s_{-i}$ is the profile of all players except $i$. Suppose, if the probability distribution $q$ is such that $q_{s^*} = 1$, then Eq.(\ref{corr eq}) reduces to the condition of Nash equilibrium for a strategy profile $s^*$. Therefore, the concept of correlated equilibrium is more general than Nash equilibrium.

For Hawk-Dove game, consider $\{q_1, q_2, q_3, q_4\}$ as the probability distribution over the strategy profiles $\{(H,H), (H,D), (D,H)\\,(D,D)\}$. Suppose, Bob deviates to $H$ from correlated strategy and Alice plays the correlated strategy, then for equilibrium condition of Eq.(\ref{corr eq}) to be satisfied,
\begin{equation}
q_1 u_B(H, H) + q_2 u_B(H, D) + q_3 u_B(D, H) + q_4 u_B(D, D) \geq q_1 u_B(H, H) + q_2 u_B(H, H) + q_3 u_B(D, H) + q_4 u_B(D, H)
\end{equation}
Where, $u_B$ maps from strategy profile to Bob's payoff. Substituting payoffs from Eq.(\ref{HDclass}) to the above inequality, and simplifying further, we get,
\begin{equation}
\frac{5}{7}q_2 - q_4 \geq 0,
\label{hden1}
\end{equation}
and if Bob deviates to $D$, then
\begin{equation}
q_1 u_B(H, H) + q_2 u_B(H, D) + q_3 u_B(D, H) + q_4 u_B(D, D) \geq q_1 u_B(H, D) + q_2 u_B(H, D) + q_3 u_B(D, D) + q_4 u_B(D, D)
\end{equation}
substituting for the payoffs we get,
\begin{equation}
\frac{5}{7}q_3 - q_1 \geq 0.
\label{hden2}
\end{equation}
Similarly, for the case where Alice alters her strategy, we get the following relations,
\begin{equation}
\frac{5}{7}q_3 - q_4 \geq 0,
\quad \text{and} \quad
\frac{5}{7}q_2 - q_1 \geq 0.
\label{hden4}
\end{equation}
Therefore, any distribution $\{q_1, q_2, q_3, q_4\}$ satisfying Eqs.(\ref{hden1}, \ref{hden2}, \ref{hden4}) with the constrain $q_1 + q_2 + q_3 + q_4 = 1$, is a correlated equilibrium.\\

For example, consider a third party which recommends the strategy profiles (H, D), (D, H) and (D, D) to the players, with probabilities 3/7, 3/7 and 1/7 respectively (i.e., $\{q_1, q_2, q_3, q_4\} = \{0, \frac{3}{7}, \frac{3}{7}, \frac{1}{7}\}$, which satisfies the inequalities). Then the expected payoff for Bob is calculated as,
\begin{equation}
\$_B = \frac{3}{7}\times u_B(H, D) + \frac{3}{7}\times u_B(D, H) + \frac{1}{7}\times u_B(D, D) = \frac{3}{7}\times 0 + \frac{3}{7}\times 50 + \frac{1}{7}\times 15 = \frac{165}{7} \approx 23.57
\end{equation}
Similarly, expected payoff for Alice is,
\begin{equation}
\$_A = \frac{3}{7}\times u_A(H, D) + \frac{3}{7}\times u_A(D, H) + \frac{1}{7}\times u_A(D, D) = \frac{3}{7}\times 50 + \frac{3}{7}\times 0 + \frac{1}{7}\times 15 = \frac{165}{7} \approx 23.57
\end{equation}

Now if Bob deviates from the correlated strategy to any other strategy, say $H$, and Alice plays from the recommended strategies then Bob's expected payoff is now,
\begin{equation}
\$_B = \frac{3}{7}\times u_B(H, H) + \frac{3}{7}\times u_B(D, H) + \frac{1}{7}\times u_B(D, H) = \frac{3}{7}\times -25 + \frac{3}{7}\times 50 + \frac{1}{7}\times 50 = 125/7 \approx 17.86
\end{equation}
Similar is the case for Alice, and yields the payoff matrix,
\begin{equation}
\begin{tabular}{|c|c|c|c|c|}
\hline
\multicolumn{2}{|c}{}& \multicolumn{3}{|c|} {\textbf{Bob}} \\
\hline
\multicolumn{2}{|c|}{} & $H$ & $D$ & $Corr$ \\
\hline
& $H$ & (-25,-25) &(50,0)&(17.86, -10.71)\\
\textbf{Alice} & $D$ & (0,50) &(15,15) &(8.57, 30)\\
& $Corr$ & (-10.71, 17.86) & (30, 8.57)&(23.57, 23.57)\\
\hline
\end{tabular},
\label{crHD1}
\end{equation}
where $"Corr"$ is the said correlated strategy, and we can see it is a correlated equilibrium yielding a more significant payoff than the mixed NE $M_{HD}$.
Another correlated equilibrium, where (H, D), (D, H) and (D, D) is suggested with probabilities
7/15, 7/15 and 1/15 yield even better payoff,
\begin{equation}
\begin{tabular}{|c|c|c|c|c|}
\hline
\multicolumn{2}{|c}{}& \multicolumn{3}{|c|} {\textbf{Bob}} \\
\hline
\multicolumn{2}{|c|}{} & $H$ & $D$ & $Corr$ \\
\hline
& $H$ & (-25,-25) &(50,0)&(15.0, -11.67)\\
\textbf{Alice} & $D$ & (0,50) &(15,15) &(8.0, 31.33)\\
& $Corr$ & (-11.67, 15.0) & (31.33, 8.0) & (24.33, 24.33)\\
\hline
\end{tabular}
\label{crHD3}
\end{equation}

And in particular, if (H, D), (D, H) is suggested with probabilities 1/2 and 1/2 respectively, then payoff approaches a maximum limit 25.
\begin{equation}
\begin{tabular}{|c|c|c|c|c|}
\hline
\multicolumn{2}{|c}{}& \multicolumn{3}{|c|} {\textbf{Bob}} \\
\hline
\multicolumn{2}{|c|}{} & $H$ & $D$ & $Corr$ \\
\hline
& $H$ & (-25,-25) &(50,0)&(12.5, -12.5)\\
\textbf{Alice} & $D$ & (0,50) &(15,15) &(7.5, 32.5)\\
& $Corr$ & (-12.5, 12.5) & (32.5, 7.5) & (25, 25)\\
\hline
\end{tabular}
\label{crHD4}
\end{equation}

Comparing Eq.(\ref{crHD4}) with mixed strategy Nash equilibrium of Eq.(\ref{mHD}), we can see that the correlated equilibrium gives way much higher payoff than mixed strategy Nash equilibrium $(M_{HD}, M_{HD})$.
}

\subsection{Pure Strategies in Quantum Hawk-Dove Game}
To introduce the quantum version of the Hawk-Dove game, we follow Marinatto and {Weber's} scheme\cite{11}. This scheme has been extended to include various forms of Hawk-Dove game with initially entangled states in Refs.\cite{12,13,14}.
The initial state $\rho_{in}$ is taken to be a maximally entangled state, $\rho_{in} = |\psi\rangle\langle\psi|$ with
\begin{equation}
|\psi\rangle = \hat{J}(\frac{\pi}{2})|00\rangle = \frac{1}{\sqrt{2}}(|00\rangle-i|11\rangle).
\end{equation}
{wherein, \begin{equation}
\hat{J}(\gamma) = \begin{pmatrix}
\cos{(\gamma/2)} & 0 & 0 & -i\sin{(\gamma/2)}\\
0 & \cos{(\gamma/2)} & i\sin{(\gamma/2)} & 0 \\
0 & i\sin{(\gamma/2)} & \cos{(\gamma/2)} & 0 \\
-i\sin{(\gamma/2)} & 0 & 0 & \cos{(\gamma/2)}
\label{Jop}
\end{pmatrix}
\end{equation} is the entanglement operator and $\gamma \in [0,\pi/2]$ is the measure of entanglement.}
The two entangled qubits are then forwarded to Alice and Bob respectively who perform unitary operation on the initial state to get-
\begin{equation}
\rho = (U_{A}\otimes U_{B}) \rho_{in} (U_{A}\otimes U_{B})^{\dagger},
\end{equation}
and then disentanglement operator $\hat{J}^{\dagger}(\frac{\pi}{2})$ is applied,
\begin{equation}
\rho_{final} = \hat{J}^{\dagger}(U_{A}\otimes U_{B}) \rho_{in} (U_{A}\otimes U_{B})^{\dagger}\hat{J}.
\end{equation}
$U$ describes a general strategy represented by a $2\times 2$ unitary matrix parametrized by two parameters $\theta \in [0,\pi]$ and $\phi \in [0,\pi/2]$, and is given as
\begin{equation}
U(\theta, \phi) = \begin{bmatrix}
e^{i\phi}\cos(\theta/2)& \sin(\theta/2)\\
-\sin(\theta/2) & e^{-i\phi}\cos(\theta/2)
\end{bmatrix}
\end{equation}
The strategy operator for the players can be formulated using this unitary operator, with Hawk ($H$) and Dove($D$) being,
\begin{equation}
H = U(\pi,0) =
\begin{bmatrix}
0&1\\
-1&0
\end{bmatrix},\quad D = U(0,0) =
\begin{bmatrix}
1&0\\
0&1
\end{bmatrix} .
\label{HD Strategy}
\end{equation}
Thus, the basis vectors of the Hilbert space will be, $|DD\rangle$ = (1, 0, 0, 0), $|DH\rangle$ = (0, 1, 0, 0), $|HD\rangle$ = (0, 0, 1, 0), $|HH\rangle$ = (0, 0, 0, 1).\\

Then the payoff operators for Alice and Bob are defined as
\begin{equation}
P_{A}= \Big(\frac{v}{2}-\frac{i}{2}\Big)|HH\rangle\langle HH| + v|HD\rangle\langle HD|+\Big(\frac{v}{2}-d\Big)|DD\rangle\langle DD|,
\end{equation}
\begin{equation}
P_{B}= \Big(\frac{v}{2}-\frac{i}{2}\Big)|HH\rangle\langle HH| + v|DH\rangle\langle DH|+\Big(\frac{v}{2}-d\Big)|DD\rangle\langle DD|.
\end{equation}
The expected payoffs for either Alice or Bob are calculated as $ \$= Tr(P\rho_{f}) $ \\

\par Now, let us consider a scenario where both the players are either restricted, or prefers to play only the quantum strategy "$Q$" \cite{6} defined as:
\begin{equation}
Q = U(0,\pi/2) =
\begin{bmatrix}
i&0\\
0&-i
\end{bmatrix},
\label{Q Strategy}
\end{equation}
and the Miracle move $"M"$ defined as:
\begin{equation}
M = U(\pi/2,\pi/2) = \frac{1}{\sqrt{2}}
\begin{bmatrix}
i&-1\\
1&-i
\end{bmatrix}.
\label{M Strategy}
\end{equation}
Then by calculating the expected payoffs using above scheme, we get the payoff table-
\begin{equation}
\begin{tabular}{|c|c|c|c|}
\hline
\multicolumn{2}{|c}{}& \multicolumn{2}{|c|} {\textbf{Bob}} \\
\hline
\multicolumn{2}{|c|}{} & $Q$ & $M$ \\
\hline
& $Q$ & (15, 15) & (32.5, 7.5)\\
\textbf{Alice} & $M$ & (7.5, 32.5) & (10, 10)\\
\hline
\end{tabular}
\label{QHD1}
\end{equation}
In this scenario, we can see that $(Q, Q)$ is pure strategy Nash equilibrium which is Pareto optimal too. It is the only dominant strategy here. There {cannot} exist a mixed strategy Nash equilibrium.
\subsection{Random strategy in quantum Hawk-Dove Game}
We now establish that using a random strategy on a maximally entangled state in a purely quantum game scenario yields a solution and payoffs which cannot be replicated in the classical Hawk-Dove game.
Let $|\psi_{in}\rangle$ be a maximally entangled state represented by:
\begin{equation}
|\psi_{in}\rangle = \hat{J}(\frac{\pi}{2})|DD\rangle = \frac{1}{\sqrt{2}}(|DD\rangle-i|HH\rangle).
\label{max ent}
\end{equation}
{
Where, $|H\rangle = \begin{pmatrix}
0\\
1
\end{pmatrix}$, $|D\rangle = \begin{pmatrix}
1\\
0
\end{pmatrix}$ and $\hat{J}$ as defined in Eq.(\ref{Jop}).} If Alice plays quantum move $Q$, with probability $p$ and the miracle move $M$ with probability
$(1 - p)$ and Bob uses these operators with probability $q$ and $(1 - q)$, respectively, then the final density matrix of the bipartite system takes the form:
$$ \rho_{f} = pq[\hat{J}^\dagger(Q_{A}\otimes Q_{B})\rho_{in}(Q_{A}^{\dagger}\otimes Q_{B}^{\dagger})\hat{J}]+ p(1-q)[\hat{J}^\dagger(Q_{A}\otimes M_{B})\rho_{in}(Q_{A}^{\dagger}\otimes M_{B}^{\dagger})\hat{J}]$$
\begin{equation}
+ (1-p)q[\hat{J}^\dagger(M_{A}\otimes Q_{B})\rho_{in}(M_{A}^{\dagger}\otimes Q_{B}^{\dagger})\hat{J}] + (1-p)(1-q)[\hat{J}^\dagger(M_{A}\otimes M_{B})\rho_{in}(M_{A}^{\dagger}\otimes M_{B}^{\dagger})\hat{J}].
\end{equation}
The expected payoff functions for both players are calculated as $ \$_{A}= Tr(P_A\rho_{f}) $ and $ \$_{B}= Tr(P_B\rho_{f})$. We will consider the random strategy as $"R_{QM}"$ where $Q$ and $M$ are played by Alice with probability $p = 1/2$ and Bob with probability $q = 1/2$. The payoff table in the larger strategic space which includes pure and the random strategy is as follows:
\begin{equation}
\begin{tabular}{|c|c|c|c|c|}
\hline
\multicolumn{2}{|c}{}& \multicolumn{3}{|c|} {\textbf{Bob}} \\
\hline
\multicolumn{2}{|c|}{} & $Q$ & $M$ & $R_{QM}$ \\
\hline
& $Q$ & (15, 15) &(32.5, 7.5)&(23.75, 11.25)\\
\textbf{Alice} & $M$ & (7.5, 32.5) &(10, 10) &(8.75, 21.25)\\
& $R_{QM}$ & (11.25, 23.75) & (21.25, 8.75)&(16.25, 16.25)\\
\hline
\end{tabular}
\label{Qhawkdove}
\end{equation}

We can see $(Q, Q)$ is still the Nash equilibrium after adding $R_{QM}$ to the payoff table. But the game is different when played with only quantum strategies with Nash equilibrium being $(15, 15)$ and Pareto optimal outcome is ($R_{QM}$, $R_{QM}$). One must also notice that the Pareto optimal $(16.25, 16.25)$ payoffs are slightly more than the Pareto optimal of the classical game. It cannot be replicated in a purely classical game, thereby negating van Enk-Pike's assertion.
\par It is also interesting to note that all the payoffs obtained in Eq.(\ref{Qhawkdove}) are greater than zero. Meaning, no players face any loss in resources when both of them {opt} to play only the said quantum strategies.

{
\subsection{Correlated equilibrium in quantum Hawk-Dove Game}
\label{corrinqhd}
Consider the restricted quantum game defined by payoff-matrix Eq.(\ref{QHD1}). Suppose $\{q_1, q_2, q_3, q_4\}$ are the probability distribution over the strategy profiles $\{(Q,Q), (Q,M), (M,Q), (M,M)\}$, and if Bob deviates to Q and Alice plays correlated strategy, then for equilibrium,
\begin{gather*}
q_1 u_B(Q, Q) + q_2 u_B(Q, M) + q_3 u_B(M, Q) + q_4 u_B(M, M) \geq q_1 u_B(Q, Q) + q_2 u_B(Q, Q) + q_3 u_B(M, Q) + q_4 u_B(M, Q)
\end{gather*}
Substituting the payoffs from Eq.(\ref{QHD1}) and simplifying we get,
\begin{equation}
7.5q_2 + 22.5q_4 \leq 0,
\label{qhden1}
\end{equation}
And similarly if Bob deviates to M and Bob remains to play the recommended strategy, then,
\begin{gather*}
q_1 u_B(Q, Q) + q_2 u_B(Q, M) + q_3 u_B(M, Q) + q_4 u_B(M, M) \geq q_1 u_B(Q, M) + q_2 u_B(Q, M) + q_3 u_B(M, M) + q_4 u_B(M, M)
\end{gather*}
substituting the payoffs we get,
\begin{equation}
7.5q_1 + 22.5q_3 \geq 0.
\label{qhden2}
\end{equation}
Similarly, for the case where Alice alters her strategy, we get the following relations,
\begin{equation}
7.5q_3 + 22.5q_4 \leq 0, \quad \text{and} \quad
7.5q_1 + 22.5q_2 \geq 0.
\label{qhden4}
\end{equation}

Since $q_2$, $q_3$, $q_4$ are positive, the inequalities of Eqs.(\ref{qhden1}, \ref{qhden4}) can only be satisfied if $q_2 = q_3 = q_4 = 0$.\\ Since, $q_1 + q_2 + q_3 + q_4 = 1$, this implies $q_1 = 1$. Therefore, the only correlated equilibrium for this restricted game is the Nash equilibrium $(Q,Q)$ itself. Here, we showed how a correlated strategy can be implemented on a discrete set of quantum strategies. \\

\subsection{Correlated games vs Quantum games}
\label{secCvsQ}
Another objection of van Enk and Pike is that entanglement in quantum games blurs the contrast between cooperative and non-cooperative games. Arguably, quantum games are a kind of correlated games constructed from independent players. A connection with Aumann's correlated games\cite{21} is quite interesting, as both players apply their respective unitary strategy operator on a maximally entangled initial state Eq.(\ref{max ent}).
van Enk and Pike's argument that entanglement violates the spirit of a non-cooperative game is vague, as clearly discussed in subsection \ref{motiv}, entanglement induces non-local correlation, and further, we perform the disentanglement operation before measurement. There is no actual binding agreement between the players as the game's entanglement ($\gamma$) is not part of the player's strategy set. In Table \ref{corr vs. quantum}, we can see how this interpretation manifests itself.
\begin{table}[h]
\begin{tabular}{|l|l|}
\hline
\multicolumn{1}{|c|}{\textbf{Correlated games}} & \multicolumn{1}{c|}{\textbf{Quantum games}} \\ \hline
\begin{tabular}[c]{@{}l@{}}- A referee or any other third party \\ implements the probability distribution \\ of the possible outcomes (strategy profiles). \\ Players have no part in forming the \\ probability distribution from which the \\ expected payoff is calculated. Still, they only have \\ the choice to play the recommended strategy \\ drawn by the referee or deviate from it.\end{tabular} & \begin{tabular}[c]{@{}l@{}}- The referee implements the measure \\ of games entanglement, and \\ the probability distribution of the \\ final state $|\psi_f\rangle$ from which the\\expected payoff is calculated is formed by the\\independent choice of player's unitary strategies \\ and the entanglement parameter.\\Players do not know if the game is entangled. \\ For the case of quantum Hawk-Dove game \\or quantum prisoner's dilemma as shown\\in Appendix, in order that the dilemma exists\\referee, should not reveal entanglement.\end{tabular} \\ \hline
\begin{tabular}[c]{@{}l@{}}- Given that a strategy has been assigned \\ to a player, from the probability distribution\\ of the strategy profile, a player can obtain \\ the conditional probabilities of the other \\ player's strategy. The player opts to play the \\ strategy assigned by the referee, supposing \\ the other player played their assigned strategy. \\ Therefore, when both the players play the \\ correlated strategy, the game is Cooperative. \end{tabular} &

\begin{tabular}[c]{@{}l@{}}- Unlike correlated games, quantum games \\do not give any information about probabilities\\ at which the other player plays his/her strategy.\\ Rather, the payoffs for a typical unitary strategy\\(e.g.,$Q$, $M$) vary with the entanglement parameter.\\Players can alter the probability distribution of the \\ final state, as players choose their unitary strategies\\ independently. Thus, refuting van Enk and Pike's\\ argument.\end{tabular} \\ \hline
\end{tabular}
\caption{A comparison between correlated games and quantum games}
\label{corr vs quantum}
\end{table}

}

\section{Discussion}
This paper explores different strategies in quantum games and how they lead to different outcomes compared to classical games. We investigate in detail the quantum Hawk-Dove game using the density matrix formalism~\cite{6} to bring out a lacuna in van Enk-Pike's assertion that equilibrium solution to a quantum game is not unique and is obtained in the underlying classical game itself. We define a random strategy "$R_{QM}$" and in order to get a unique solution of the quantum game that cannot be replicated in the classical game.
Thus, the essence of a quantum game can never be completely captured by the original classical game.
\par Recently, in support of van Enk-Pike's assertion, Ref.~\cite{15} argues the following points,
\begin{enumerate}
\item The quantum game is a different game, and solving this game doesn't amount to solving the original classical game.
\item The payoffs in a quantum game are all obtained in a classical game.
\end{enumerate}
We agree on the first point but, the second point is correct only for the pure quantum strategy $"Q"$, and it isn't accurate if we change to Miracle move strategy $"M"$ or random strategy $"R_{QM}"$. The second point is what we contest. Further, we showed that payoffs corresponding to a random quantum strategy $R_{QM}$ in a quantum game could not be obtained in a classical game. We have also shown that, in a purely quantum game, where players prefer to play only quantum strategy, the random quantum strategy gives rise to a new and unique equilibrium. Especially, we have shown that a purely quantum Hawk-Dove game gives rise to one dominant equilibrium strategy and a new Pareto optimal that is higher than Pareto optimality in a purely classical game. On the other hand, Ref.~\cite{16} has shown that introducing a mixed random strategy won't violate the rules of the game\cite{16} agreeing with our contention.

\par The emphasis of this work was on examining how a random strategy applied in the context of quantum Hawk-Dove game provides a unique solution than what is possible in the classical equivalent. Our work leads us to conclude that we can indeed get exclusive solutions to quantum games which aren't possible classically, by using proper strategies negating the conclusions of van Enk-Pike. Strategies control the outcome of any game. We have to choose the strategy properly to get the best solution to a particular game. We hope our results may enable everyone to understand better the structure of quantum games and their application in quantum information theory.
\section{Acknowledgment}
CB would like to thank Science and Engineering Research Board (SERB) for funding under MATRICS grant ''Nash equilibrium versus Pareto optimality in N-Player games`` (MTR/2018/000070).

\appendix
\section{Random mixed strategies in quantum Prisoner's Dilemma}
A similar calculations as done for quantum Hawk-Dove can be done for Prisoner's dilemma. The classical payoff table for pure strategies in PD is represented as-
\begin{equation}
\begin{tabular}{|c|c|c|c|}
\hline
\multicolumn{2}{|c}{}& \multicolumn{2}{|c|} {\textbf{Bob}} \\
\hline
\multicolumn{2}{|c|}{} & $C$ & $D$ \\
\hline
& $C$ & $(3,3)$ &$(0,5)$\\
\textbf{Alice} & $D$ & $(5,0)$&$(1,1)$\\
\hline
\end{tabular}
\label{pda1}
\end{equation}
Where "$C$" and "$D$" represent the strategies corresponding to pure \textit{Confess(C)} and \textit{Defect(D)}. The Nash equilibrium is (D, D) here.

Since, (D, D) is the only one dominant strategy in PD, there can't exists a Mixed strategy Nash equilibrium. But we may expand the strategic space with a Random strategy $"R_{CD}"$ where, p = q = 1/2
\begin{equation}
\begin{tabular}{|c|c|c|c|c|}
\hline
\multicolumn{2}{|c}{}& \multicolumn{3}{|c|} {\textbf{Bob}} \\
\hline
\multicolumn{2}{|c|}{} & $C$ & $D$ & $R_{CD}$ \\
\hline
& $C$ & (3,3) &(0,5)&(1.5, 4)\\
\textbf{Alice} & $D$ & (5,0) &(1,1) &(3, 0.5)\\
& $R_{CD}$ & (4, 1.5) & (0.5, 3)&(2.25, 2.25)\\
\hline
\end{tabular}
\end{equation}
$(D, D)$ is still the Nash equilibrium after a adding mixed strategy to the payoff table.\\

When Alice and Bob are restricted to play only the Quantum Strategies $Q$ and $M$ defined in equation(22-23) in prisoners dilemma, the following payoff matrix can be calculated using the Marinatto and {Weber's} scheme\cite{11} we incorporated in section II (by taking $C$ and $D$ as $U(0,0)$ and $U(\pi,0)$ this time).

\begin{equation}
\begin{tabular}{|c|c|c|c|}
\hline
\multicolumn{2}{|c}{}& \multicolumn{2}{|c|} {\textbf{Bob}} \\
\hline
\multicolumn{2}{|c|}{} & $Q$ & $M$ \\
\hline
& $Q$ & (3, 3) & (4, 1.5)\\
\textbf{Alice} & $M$ & (1.5, 4) & (2.25, 2.25)\\
\hline
\end{tabular}
\label{pda3}
\end{equation}

In this scenario, we can see that (Q, Q) becomes the Nash equilibrium and Pareto optimal. And there can't exist a mixed strategy Nash equilibrium since there is only one dominant strategy.

Now let $|\psi_{in}\rangle$ be a maximally entangled state represented by:
\begin{equation}
|\psi_{in}\rangle = \hat{J}|00\rangle = \frac{1}{\sqrt{2}}(|00\rangle-i|11\rangle).
\end{equation}
If Alice uses $Q$, the quantum strategy, with probability $p$ and $M$ with probability
$(1 - p)$ and Bob uses these operators with probability $q$ and $(1 - q)$, respectively.
Then the final density matrix of the bipartite system takes the form:
$$ \rho_{f} = pq[\hat{J}^\dagger(Q_{A}\otimes Q_{B})\rho_{in}(Q_{A}^{\dagger}\otimes Q_{B}^{\dagger})\hat{J}]+ p(1-q)[\hat{J}^\dagger(Q_{A}\otimes M_{B})\rho_{in}(Q_{A}^{\dagger}\otimes M_{B}^{\dagger})\hat{J}]$$
\begin{equation}
+ (1-p)q[\hat{J}^\dagger(M_{A}\otimes Q_{B})\rho_{in}(M_{A}^{\dagger}\otimes Q_{B}^{\dagger})\hat{J}] + (1-p)(1-q)[\hat{J}^\dagger(M_{A}\otimes M_{B})\rho_{in}(M_{A}^{\dagger}\otimes M_{B}^{\dagger})\hat{J}]
\end{equation}
Here $\rho_{in}=|\psi_{in}\rangle\langle\psi_{in}|$. The payoff operators for Alice and Bob are defined as
\begin{equation}
P_{A}= 3|00\rangle\langle 00| + 5|10\rangle\langle 10|+|11\rangle\langle 11|,
\end{equation}
\begin{equation}
P_{B}= 3|00\rangle\langle 00| + 5|01\rangle\langle 01|+|11\rangle\langle 11|.
\end{equation}
The payoff functions for Alice and Bob are the mean values of the above operators, i.e.,

\begin{equation}
\$_{A}(p,q)= Tr(P_{A}\rho_{f}) \hspace{0.5cm} and \hspace{0.5cm} \$_{B}(p,q)= Tr(P_{B}\rho_{f}).
\end{equation}

Quantum payoff table with random strategy is:
\begin{equation}
\begin{tabular}{|c|c|c|c|c|}
\hline
\multicolumn{2}{|c}{}& \multicolumn{3}{|c|} {\textbf{Bob}} \\
\hline
\multicolumn{2}{|c|}{} & $Q$ & $M$ & $R_{QM}$ \\
\hline
& $Q$ & (3,3) &(4, 1.5)&(3.5, 2.25)\\
\textbf{Alice} & $M$ & (1.5, 4) &(2.25, 2.25) &(1.88, 3.12)\\
& $R_{QM}$ & (2.25, 3.5) & (3.12, 1.88)&(2.69, 2.69)\\
\hline
\end{tabular}
\end{equation}

In this case, clearly, $(Q, Q)$ is NE and Pareto optimal with an added mixed quantum strategy in the payoff table. But the payoff obtained here too cannot be replicated in a pure classical game, negating van Enk-Pike's assertion again in a quantized prisoner's dilemma too. \\

{
Via a similar approach used to calculate correlated equilibrium of quantum Hawk-Dove game in subsection\ref{corrinqhd}, it can be shown that the only correlated equilibrium for classical Prisoner's dilemma(\ref{pda1}) and the quantum prisoner's dilemma(\ref{pda3}) is the Nash equilibrium of the game itself.
}

\end{document}